\documentstyle[epsfig]{elsart}
\begin{document}
\begin{frontmatter}
\title{A measurement of the absolute neutron beam polarization
produced by an optically-pumped $^{3}$He neutron spin filter}
\author[IUCF]{D.~R.~Rich},
\author[LANL]{J.~D.~Bowman},
\author[TUNL]{B.~E.~Crawford\thanksref{get}},
\author[TRIUMF]{P.~P.~J.~Delheij},
\author[UMN]{M.~A.~Espy\thanksref{lanl}},
\author[KYOTO]{T.~Haseyama},
\author[NIST]{G.~Jones\thanksref{ham}},
\author[IUCF]{C.~D.~Keith\thanksref{cebaf}},
\author[LANL]{J.~Knudson},
\author[UNH]{M.~B.~Leuschner},
\author[KYOTO]{A.~Masaike\thanksref{fukui}},
\author[KEK]{Y.~Masuda},
\author[KYOTO]{Y.~Matsuda\thanksref{riken}},
\author[LANL]{S.~I.~Penttil\"a},
\author[UNH]{V.~R.~Pomeroy},
\author[LANL]{D.~A.~Smith},
\author[IUCF]{W. M.~Snow},
\author[NCSU]{S.~L.~Stephenson\thanksref{get}},
\author[NIST]{A.~K.~Thompson} and
\author[LANL]{V.~Yuan}
\address[IUCF]{Indiana University, Bloomington, IN 47405}
\address[LANL]{Los Alamos National Laboratory, Los Alamos, NM 87545}
\address[TUNL]{Duke University, Durham, NC 27708 and Triangle
Universities Nuclear Laboratory, Durham, NC 27708}
\address[TRIUMF]{TRIUMF, Vancouver, BC, Canada V6T 2A3}
\address[UMN]{University of Minnesota, Minneapolis, MN 55455}
\address[KYOTO]{Kyoto University, Kyoto 606-01, Japan}
\address[NIST]{NIST, Gaithersburg, MD 20899}
\address[UNH]{University of New Hampshire, Durham, NH 03824}
\address[KEK]{Laboratory for High Energy Physics, 1-1, Oho, Tsukuba 305,
Japan}
\address[NCSU]{North Carolina State University, Raleigh, NC 27695 and
Triangle 
Universities Nuclear Laboratory, Durham, NC 27708}
\thanks[lanl]{Present address: Los Alamos National Laboratory, Los Alamos,
NM 
87545} 
\thanks[ham]{Present address: Hamilton College, Clinton, NY 13323}
\thanks[cebaf]{Present address: Thomas Jefferson National
Accelerator Facility, Newport News, VA 23606}
\thanks[fukui]{Present address: Fukui University of Technology, 3-6-1
Gakuen, 
Fukui-shi, Japan}
\thanks[riken]{Present address: Institute of Physical and Chemical
Research 
(RIKEN), Saitama, 351-8526, Japan}
\thanks[get]{Present address: Gettysburg College, Gettysburg, PA 17325}

\begin{abstract}
The capability of performing accurate absolute measurements of neutron beam
polarization opens a number of exciting opportunities in fundamental
neutron physics and in neutron scattering.  At the LANSCE pulsed neutron
source we have measured the neutron beam polarization with an absolute
accuracy of 0.3\% in the neutron energy range from 40~meV to 10~eV
using an optically-pumped polarized $^{3}$He spin filter and a relative
transmission measurement technique. $^3$He was polarized using the Rb
spin-exchange method. We describe the measurement technique, present our
results, and discuss some of the systematic effects associated with the
method. 



\end{abstract}
\end{frontmatter}
\section{Introduction}
Modern experiments using low-energy neutrons address issues of importance
in nuclear, particle, and astrophysics~\cite{dub91}.  In particular,
precision measurements of neutron decay parameters such as the decay rate
and angular correlation coefficients of the decay products are of
fundamental importance. In combination with a separate measurement of the
neutron decay rate, for example, a measurement of the electron asymmetry
coefficient in polarized neutron decay (the $A$ coefficient) can be
used to determine the weak polar vector-axial vector coupling ratio
$\lambda={g_{A}\over g_{V}}$ and, by comparison with muon decay, the
Kobayashi-Maskawa (KM)
matrix element $V_{ud}$, one of the parameters of the Standard Model of
elementary particle interactions~\cite{fre90,abe97}. 

The hypothesis that the KM matrix is
unitary implies that the sum of the squares of each row are unity. The
uncertainty of the unitarity test is dominated by the accuracy of the
large diagonal elements $V_{ud}$, $V_{cs}$, and $V_{tb}$.  While at
present, the best limits on the unitarity of the KM matrix are extracted
from measurements of superallowed beta decay, in the long run study
of neutron beta decay promises to provide even better
limits~\cite{hard}.
While it is difficult to measure $V_{ud}$ to $0.1\%$,
there is no hope of measuring $V_{cs}$ or $V_{tb}$ to this accuracy in the
foreseeable future.  It is important to note
that present measurements in the neutron sector are marginally
inconsistent with the unitarity of the KM
matrix~\cite{car90,th}: the value of $V_{ud}$ inferred from the $A$
measurements of comparable accuracy corresponds to a violation of the
unitarity of the KM matrix by about $3\sigma$.  

Also, measurements of the neutrino asymmetry $B$ in
polarized neutron decay can be used to place interesting constraints on
possible deviations from the Standard Model in the charged current sector
of the weak interaction, such as the possible existence of right-handed
weak currents~\cite{aps98}. It is therefore important to improve the
accuracy of these measurements. 

Unfortunately, both of the most accurate measurements of the $A$ and $B$
coefficients in neutron decay suffer from a common limitation: the
absolute accuracy of the neutron beam polarization measurement. To
significantly improve the measurements alternative techniques are
needed. One
approach is to conduct the measurement with very low energy neutrons
(known as ultracold neutrons, see~\cite{gol91} for an introduction).
Ultracold neutrons (UCNs) can be prepared in a definite state of
polarization relative to an external magnetic field by passing them
through a region in which the $\vec{\mu}\cdot\vec{B}$ interaction is
larger than the kinetic energy for one of the spin states.  UCN
production rates are currently limited.

Another strategy, and the one pursued in this work, is to develop more
accurate methods of measuring the
polarization of neutron beams.  In order to make significant progress, a
method of absolute neutron beam polarization measurement with an accuracy
better than 0.1\% is required.

Past techniques for absolute neutron beam polarization measurement in the
cold, thermal, and epithermal neutron energy ranges fall into two main
classes:  (1) relative intensity measurements of the spin components in a
polarized beam  after spatial separation using a magnetic field gradient
(the Stern-Gerlach effect)~\cite{she54,has62,ero64,bar68,ham75} and (2)
transmission measurements using a second polarizing device (an analyzer)
in combination with spin
flippers~\cite{hal41,hug48,hug51,shu51,ken69,vor77,nas91,ero64,ser95}. 
In principle, both techniques can be implemented in
a manner which is free of systematic errors for
monoenergetic, divergence-free neutron beams.  In practice, however, there
are limitations to both of these techniques that are associated with the
phase space density of realistic neutron beams. A detailed discussion of
some of the difficulties that are encountered with these techniques has
appeared recently~\cite{yer99}.

The use of polarized $^3$He gas targets as neutron polarizers makes
possible a promising technique for absolute neutron beam polarization
measurement.  For neutrons of a given energy, there is a simple (and
essentially exact) relation between
the neutron polarization $P_n$ produced by transmission through a
polarized target and the relative transmission of neutrons through the
polarized ($T_{n}$) and unpolarized ($T^{0}_{n}$) target: 
\begin{equation}
P_n = \sqrt{1-(T^{0}_{n}/T_{n})^2}.
\label{eq:polVratio}
\end{equation}
$^3$He is an attractive target for the development of this technique.
Low energy (thermal) neutrons interact with $^{3}$He primarily by the
absorption reaction $n$~+~$^{3}He$~$\to$~$^{3}H$~+~$p$ with a cross section
$\sigma_0=5333 \pm 7$ barns at 25.3 meV~\cite{als64}. 
All other channels have comparatively small cross sections at thermal
energies.  A $^{3}$He target that is thick for absorption is thin for
the other reaction channels and the equation above holds to a high degree
of accuracy under realistic beam conditions and detector geometries. 
Since this relation requires only relative neutron transmission
measurements, it promises to form the basis of a new and reliable way to
measure and monitor
on-line the absolute polarization of a neutron beam \cite{gre95}. 

The realization of this method of neutron beam polarization measurement
has been made possible by recent advancements in techniques to polarize
$^3$He gas by optical pumping. Two methods have been developed to
polarize ground state $^3$He gas atoms: spin exchange with optically
polarized Rb~\cite{chu95,wal97}, and optical pumping of metastable $^3$He
followed by metastability exchange collisions.  Both
techniques have been applied at the ILL in Grenoble and at NIST to
polarize neutron beams~\cite{tas92,sur97,gen99}.  The Rb spin exchange
method has been used to produce polarized $^3$He targets for several
particle and nuclear physics experiments~\cite{lar91,chu92,joh95,esp98}.
The first experiment using a $^3$He spin filter to polarize an epithermal
neutron beam was performed at Los Alamos by Coulter {\it et al.} in
1989~\cite{cou90}.  The neutron beam polarization was determined in this
initial experiment with an uncertainty of 10\%.  They used a dye laser to
polarize Rb atoms.  Recent progress in the Rb spin exchange technique has
been aided largely by the commercial availability of high power laser diode
arrays~\cite{cum95}. 

The technique described in Ref.~\cite{gre95} assumed that the neutron
beams to be polarized were from a CW source with a broad energy spectrum,
and thus required a
set of transmission measurements through a reference cell to effectively
measure the energy spectrum.  At a short
pulsed spallation neutron source the energy of the neutrons can be
determined very accurately using time-of-flight (TOF) techniques.  In
addition, the pulsed nature of the source has other advantages in the
determination of detector backgrounds and {\it in situ} control of
systematic errors (as discussed further in section 4). 

In this paper, we describe the first accurate absolute measurement of neutron 
beam polarization using this technique. Our measurement,
performed at the LANSCE pulsed neutron source at Los Alamos, has achieved
an absolute accuracy of 0.3\% in the neutron energy range of 40 meV to 10
eV.  This accuracy is comparable to the best previous measurements using
supermirror neutron polarizers at energies of a few meV, and to our
knowledge represents the most accurate absolute measurement of neutron
beam polarization made in this energy range.  We argue that
the absolute accuracy can be improved by about an order of magnitude and that 
the technique can be extended to lower neutron energies.  This development
therefore promises to remove one of the
obstacles to the goal of improving the accuracy of measurements of the $A$
and $B$ coefficients in neutron beta decay.  In 
addition, the result is important for the use of polarized $^3$He spin 
filters as neutron polarizers and polarization analyzers. 

The remainder of the paper is organized as follows. Transmission of
unpolarized neutrons through polarized gas is discussed in section 2.
Section 3 discusses the LANSCE pulsed neutron source and
experimental setup with the $^3$He spin filter.
Data analysis and results are presented in section 4
and discussed in section 5.

Finally, we note that another approach which exploits the properties of
polarized $^3$He is being pursued in which $^3$He is used as an ideal
neutron polarization analyzer~\cite{zimm}.
\section{Neutron Transmission through Polarized $^3$He Gas}
Polarized $^3$He can be used as a nearly perfect neutron-spin filter
because of its very large, spin-dependent neutron reaction cross section. In 
this section we review what is known regarding the interactions of low energy 
neutrons with $^{3}$He and describe the operation of a neutron spin filter.

We begin by separating the interaction of low energy neutrons with
$^{3}$He into elastic and inelastic channels.  We first identify the
following contributions to the elastic
cross section: (1) potential scattering of the neutron from the $^{3}$He
nucleus due to the strong  interaction, (2) the electromagnetic spin-orbit
interaction of the magnetic moment of the moving neutron with the
magnetic fields originating from the electric charge of the nucleus and
bound electrons, (3) the electromagnetic interaction of the internal
charge distribution of the neutron with the bound electrons and nucleus of
the $^{3}$He atom, and (4) the electromagnetic interaction of the induced
electric dipole moment of the neutron with the electric field of the
nucleus. In the energy range of these measurements, the 
relative sizes of the scattering lengths for these processes are
approximately $4:3\times 10^{-3}:10^{-3}:10^{-4}$. Clearly (1)
dominates.  The potential scattering $n~+~^3He~\to~n~+~^3He$ is
independent of neutron energy in this energy range and can
be spin dependent in principle.  However, the elastic cross section is 
small compared to the inelastic cross section in our energy range and we
will not need to know anything about the possible spin dependence of this
cross section.  The electromagnetic spin-orbit scattering, which is spin
dependent, vanishes for forward scattering and is therefore also
negligible.

In addition to elastic scattering, there are two inelastic channels 
available for neutrons on $^{3}$He. 
\begin{equation}
n + {^3He} \to {^3H} + p ; n+{^3He} \to {^4He} + \gamma
\label{eq:inelastic}
\end{equation}
The energy dependence of both cross sections
obeys the well known $1/v$ law:  $\sigma_{re} = \frac{v_0}{v}\sigma_0$,
where $\sigma_0$ is the spin averaged cross section for neutrons moving at
a speed of $v_0$ = 2200 m/s, which corresponds to a neutron energy of 25.3
meV~\cite{als64}.  The reaction $n+{^3}He \to {^4}He+\gamma$
has a cross section of 54 $\mu$barns at 25.3 meV~\cite{wol89} and will be
ignored in the remaining discussion.  Due to the presence of a broad (400
keV) $J^{\pi} = 0^+$ excited state of the $^4$He compound nucleus, located
650 keV below the $n + {^3}He$ threshold, the reaction
$n~+~^{3}He~\to~^{3}H~+~p$ has a very large cross section $\sigma_0$ of
5333$\pm$7 barns at this energy~\cite{als64}.  Since this resonance is
only open in the $0^+$ channel, the absorption cross section is quite spin
dependent.  The energy dependence of
this reaction has been measured to high accuracy in this energy range, 
and has been shown to obey the $1/v$ law~\cite{kei98}.  

To understand the spin dependence of this reaction, we consider the
following formalism.  We consider the low energy regime where only $l = 0$
partial waves contribute (the $l = 1$ contribution to the cross section at
neutron energies $\leq$ 10~eV is at the $10^{-4}$ level). For
incident neutrons on a spin 1/2 target, we define $\sigma_s$
and $\sigma_t$ as the cross sections for the singlet and triplet compound
states, respectively.  The spin independent cross section is then:
\[\sigma_{re} = \frac{I+1}{2I+1}\sigma_t + \frac{I}{2I+1}\sigma_s\]
and the cross section for a particular spin state (the polarized cross
section) is:
\[\sigma_p = \frac{I}{2I+1}(\sigma_s - \sigma_t)\]
where $I$ is the target spin.  The ``experimental'' cross sections for
neutron and target spins parallel (+) or antiparallel (--) can be defined:
\[\sigma_{\pm} = \sigma_{re} \mp P_{N}\sigma_p\]
where $P_N$ is the target polarization.  For $I = 1/2$ and $P_N$ =
100\%:
\[\sigma_{re} = (3/4)\sigma_t + (1/4)\sigma_s\]
\[\sigma_p = (1/4)(\sigma_s - \sigma_t)\]
\[\sigma_+ = \sigma_t\]
\[\sigma_- = (\sigma_s + \sigma_t)/2\]
For the case of $^3$He, Passell and Schermer~\cite{ps66} have measured
$\sigma_p/\sigma_{re}$ = 1.010(32), making:
\[\sigma_+ = \sigma_t = 0\]
\[\sigma_- = 1/2\sigma_s = 2\sigma_{re}\]
Polarized $^3$He targets with 100\% polarization are therefore ideal
neutron polarizers.

The transmission for unpolarized neutrons with spins parallel and
antiparallel to the $^3$He spin, on an ensemble of
polarized $^3$He nuclei with polarization $P_{He}$ is given by
\begin{equation}
t_{\pm} = {\rm exp}[-nl(\sigma_{re} \mp \sigma_{p}P_{He})],
\label{eq:trans1}
\end{equation}
where $n$ is the number density and $l$ is the length of the sample.  

Polarization of the transmitted beam is then
\begin{equation}
P_n = (t_+ - t_-)/(t_+ + t_-) = {\rm tanh}(nl\sigma_{p} P_{\rm He})
\label{eq:pol1}
\end{equation}
and the transmission through the filter is given by
\begin{equation}
T_{n} = (t_+ + t_-)/2 = {\rm exp}(-nl \sigma_{re}){\rm cosh}(nl
\sigma_{p} P_{\rm He}) =T^{0}_{n}{\rm cosh}(nl \sigma_{p} P_{\rm He}).
\label{eq:trans2}
\end{equation}
Here $T_n^0$ is the transmission through the $^3$He cell when
the $^3$He polarization is zero.  Because the absorption cross section is
inversely proportional to the neutron velocity, and the scattering cross
section is nearly constant over the energy range of interest, this
relation has a simple energy dependence which we exploit in the data
analysis.

For the beam polarization we obtain equation~\ref{eq:polVratio} from
equations \ref{eq:pol1} and \ref{eq:trans2}
\[P_n = \sqrt{1-(T^{0}_{n}/T_{n})^2}.\]
According to this result, we need only measure the transmission
ratio $(T^{0}_{n}/T_{n})$ as a function of neutron energy
$E_{n}$ caused by $^3$He polarization to determine the neutron beam
polarization at $E_n$. We do not need to know the areal density of the
$^3$He, the cross sections, or the $^3$He polarization. Only an accurate
relative transmission measurement is required.

\begin{figure}
\scalebox{1.2}{
\includegraphics*[1.5in,4.5in][8.25in,8.8in]{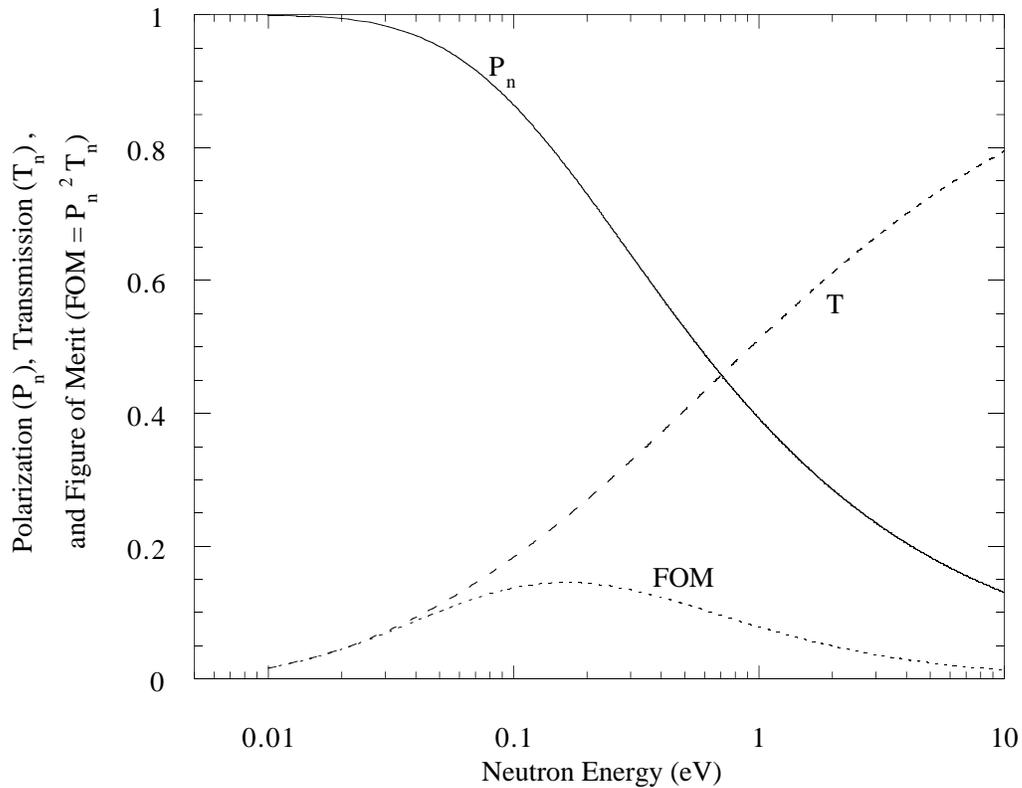}}
\caption                
{
A neutron polarization, $P_{\rm n}$ (solid), transmission, $T_{\rm n}$
(dashed) and figure-of-merit (FOM), $P_{\rm n}^{\rm 2}T_{\rm n}$ (dotted)
for a neutron spin filter with 55\% $^{3}$He polarization and thickness of
30~bar-cm as a function of neutron energy.
}
\label{fig:char}
\end{figure}

If the polarized $^3$He cell is used to analyze the neutron spin
then the analyzing power is
\begin{equation}
A=P_n {\rm tanh}(nl \sigma_{p} P_{\rm He}).
\label{eq:analyze}
\end{equation}
Figure~\ref{fig:char} shows characteristics of a $^3$He spin filter for a
cell with $^3$He thickness of 30 bar-cm (corresponding to an areal
density of $1.5 \times 10^{21}$ cm$^{-2}$) and $^3$He
polarization of 55\%. In
this figure, the neutron beam polarization,
transmission and figure-of-merit (FOM)=$P_n^2T_n$ are plotted as a
function of neutron energy. The FOM is inversely proportional to
the running time of experiments whose error is dominated by
the counting statistics of a signal depending on neutron polarization
\cite{tas95}.

Figure~\ref{fig:char} demonstrates the useful energy range
of a spin filter with a fixed $^3$He thickness.
To date the reported $^3$He polarizations with the rubidium spin-exchange
method in large-volume cells are in the range of 40\%--70\%
\cite{cum95,lar91,joh95}.  It is then possible to build practical $^3$He
neutron spin filters for neutron energies up to tens of eV, limited
ultimately by the $1/v$ velocity dependence of the neutron absorption
cross section.  The energy range of supermirror polarizers
\cite{sch89,mez76}
extends from cold to thermal neutron energies, but at epithermal energies
the $^3$He spin filter competes only with the cryogenic polarized proton
spin filter \cite{pen95}, which operates by spin-dependent scattering as
opposed to spin-dependent absorption.
\section{Accurate Neutron Beam Polarization Measurement}
In this section we describe in detail the aspects of the neutron beamline
and experimental apparatus which are important for understanding the 
measurements described in section 4. 
\subsection{LANSCE Pulsed Spallation Neutron Source}
At the Manuel Lujan Neutron Scattering Center (MLNSC) at LANSCE,
neutrons are produced by spallation.  First, 500~$\mu$s-wide, 800-MeV
proton pulses are injected into the Los Alamos Proton Storage Ring (PSR).
Inside the PSR, the pulses are compressed into triangular pulses 250~ns
wide at the base before delivery to the spallation target
at a rate of 20~Hz, producing unpolarized neutrons in the MeV energy
range.  The spallation target consists of two tungsten cylinders of 20~cm 
diameter.  The cylinders are located
one above the other along the vertical proton beam
        axis.  The upper cylinder has a length of 7~cm, while
        the lower one is 27~cm long.  The gap between them
        is 14~cm and is surrounded by moderation and
        reflector material.  The horizontal flight paths are 
        located at the level of the gap
        to minimize contamination of the beam
        from gamma-rays and high-energy neutrons.  The experiment
        described here was performed on a flight path with a
        gadolinium-poisoned water moderator and a
        cadmium/boron liner. A  more detailed description of
the spallation source and the associated beamline is given in 
\cite{rob93,lis90}.  

\begin{figure}
\scalebox{1.2}{
\includegraphics*[1.6in,4.25in][8in,8in]{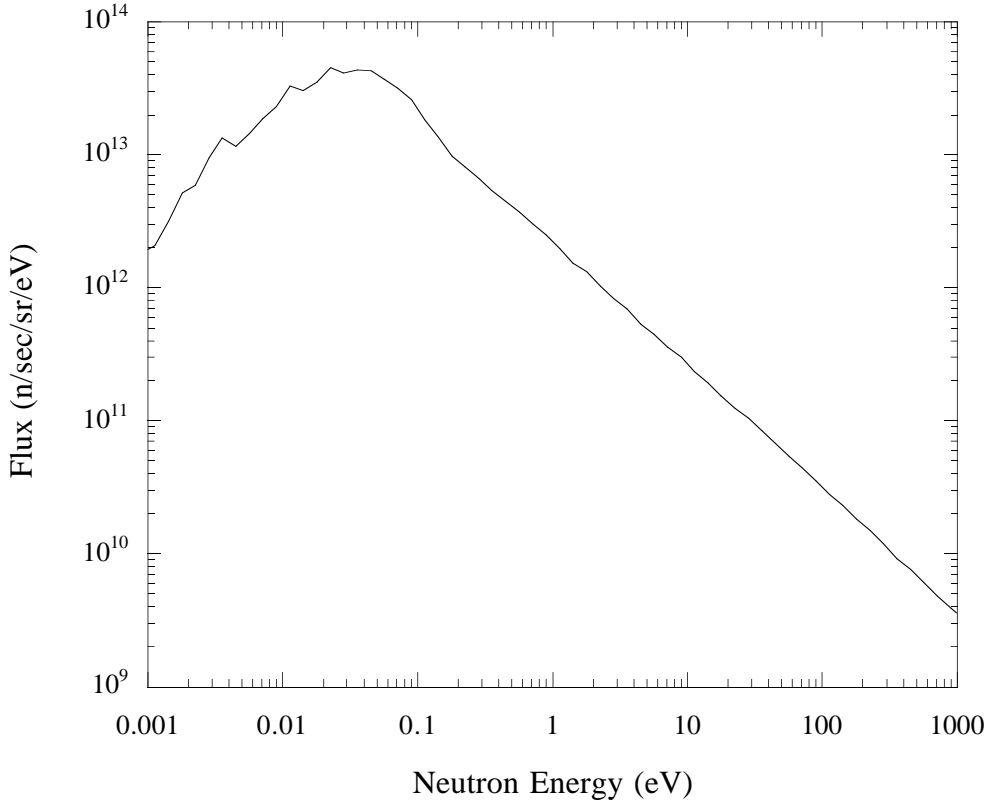}}
\caption
{
Calculated neutron flux from the water moderator at MLNSC. This is the
total neutron flux from the 12.5~cm$\times$12.5~cm surface of the
high resolution water moderator
with an average proton current of $70~\mu$A.
}
\label{fig:beamflux}
\end{figure}

The neutron yield from the
water moderator surface has an approximately Maxwell-Boltzmann
distribution at the effective temperature of the moderator, with
a high-energy tail that falls off approximately like $1/E_n$. This is
shown in figure~\ref{fig:beamflux}, which displays the neutron flux on the
surface of the moderator calculated for an average proton current of
$70~\mu$A.  The peak of the neutron flux for the water moderator used
in this experiment, $4 \times 10^{13}$ neutrons/(eV$\cdot$s$\cdot$sr), is
at about 40~meV~\cite{lis90}. $\Delta N$, the number of neutrons per
second in the epithermal energy range with energies between $E$ and
$E+\Delta E$ in the beam, is approximately given by
\begin{equation}
\Delta N =N_{0} \frac{\Delta E}{E^{0.96}}f\Omega .
\label{eq:flux}
\end{equation}
Here $N_{0}=\frac{2 \times 10^{12}}{2\pi}~{\rm neutrons}/({\rm s \cdot \rm
sr})$, $f$ is the fraction of the $12.5\times 12.5$~cm$^2$ moderator
surface the detector views through the collimation, and $\Omega$ is
the detector solid angle ($f\Omega=5.5\times 10^{-7}$~sr).

The accuracy of the time-of-flight measurement depends on the
length of the flight path (which in this experiment was about 60~m)  and
the width of the proton pulse.
Further  time broadening of the neutron pulse is introduced by the neutron
moderation processes and the fact that the normal of the moderator
surface is at a $15$ degree angle with respect to the flight path, leading
to a distribution of lengths of the flight path of the neutrons across the
beam profile.  When the neutrons interact with the moderator,
the neutron time spread can be described by the sum of two convolutions
between a Gaussian and a pair of exponential tails as
discussed in Ref.~\cite{cra97}. These time broadenings have an energy
dependence.  All of these effects are taken into account by the code FITXS 
\cite{bow97} used to fit the energy calibration data.
\subsection{Experimental Setup}
\begin{figure}
\includegraphics*[1.25in,1.75in][7.5in,8in]{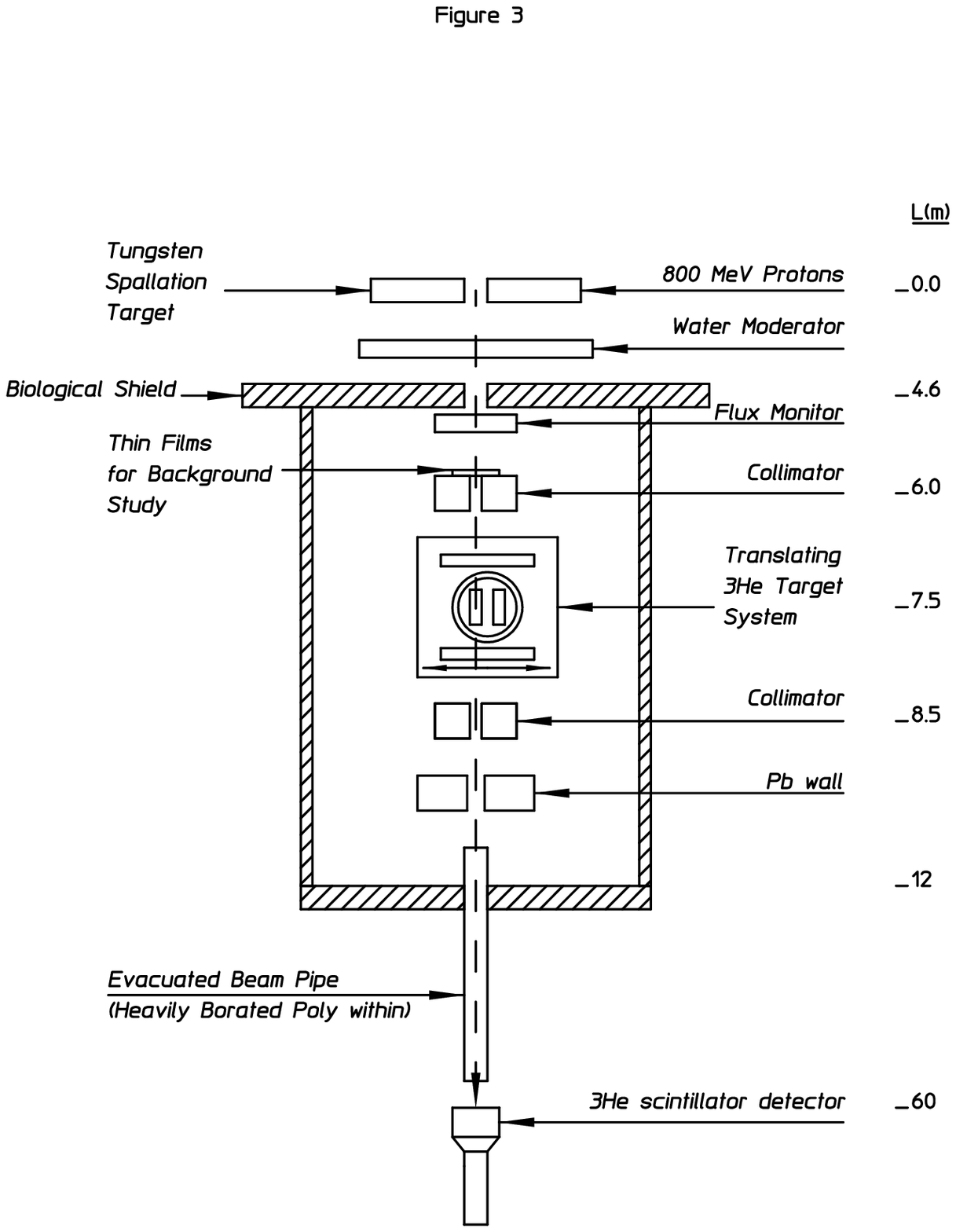}
\caption
{
Conceptual layout of the beam polarization measurement (not to scale).
The most important elements and their
approximate distances from the source are indicated.
}
\label{fig:beamscheme}
\end{figure}

The goal of the experiment was to study polarized
$^3$He as a neutron spin filter, determine the accuracy of the
beam polarization measurement technique, and discover the important
sources of systematic error. Figure \ref{fig:beamscheme} shows
schematically the main components of our experimental set up in flight
path 2 at MLNSC.

A thin $^3$He/$^4$He ion chamber combination was 
used to normalize the beam flux, with a statistical
accuracy of $10^{-4}$ per neutron pulse~\cite{zym90}.
After the beam monitor, the neutrons were collimated to a beam of 1.1 cm
in diameter. Then the neutrons interacted with a cell of $^3$He, located 9 m 
from the source.  The $^3$He apparatus was mounted on a movable table so
that two $^3$He cells, one polarized and one unpolarized,
could be alternated in and out of the beam. 

In addition to neutrons the
beam contains a TOF-dependent gamma ray background.
Detector backgrounds were determined by a modification of the absorber
techniques described in Ref.~\cite{yen97}.
A 0.13 mm thick In foil and a 2.6 mm thick Ta foil were both present in
the beam during the experiment.  The resonance absorption in
these nuclei was used to determine the gamma background
on-line --- at energies corresponding to compound nuclear resonances in
these nuclei, the neutron absorption cross section is large enough that,
for these thicknesses, essentially no neutrons pass through the absorbers.

Because of  the Maxwellian energy
distribution of neutrons from the moderator, a few very low-energy
neutrons are present. If detected, these neutrons can overlap in time with
the faster neutrons of the next pulse (\lq\lq frame overlap\rq\rq) and 
complicate the interpretation of the transmission data.  In our measurements, 
these low energy neutrons are absorbed in the $^3$He targets, due to the
very large cross section of neutrons on $^3$He at these low energies.  The
water moderator is also poisoned with Gd to absorb low energy neutrons. As
a result we expect (and observed) no frame overlap in our measurements.

Collimation and shielding
of the neutron beam is accomplished by a
        combination of brass, lead, and polyethylene loaded
        with either boron or lithium.  Brass collimators located
        immediately up-stream and downstream of the $^3$He gas cell define
        a 1.1~cm diameter beam.  A lead wall with a 10~cm square
        hole is located downstream of the second set of brass
        collimators and limits the flux of fast neutrons entering
        the 40~m long evacuated beam pipe leading to
        the $^3$He scintillation detector.  A series
        of borated polyethylene collimators with 10~cm
        diameter holes are located at regular intervals
        inside the beam pipe.  A wall of lithium-loaded 
        polyethylene with a 10~cm square hole is constructed
        3~m from the detector.  Attempts to locate shielding tightly
        around the detector were found to increase the background.
Therefore the detector was left unshielded.

Neutrons were detected with a $^3$He gas scintillation detector. 
Discriminated PMT pulses were counted with a multiscaler.  The neutron
detector and associated electronics are described in section 3.4.
\subsection{$^3$He Spin Filter}
\subsubsection{Polarizing $^3$He with the Rb spin-exchange}
$^3$He gas was polarized using the Rb-spin
exchange method.  Intense circularly-polarized laser light ($\lambda
\approx 795$ nm) is used to optically pump the $D1$ line of Rb atoms,
polarizing them, in a glass sample cell.  During binary collisions between
the $^3$He and atomically polarized Rb in the cell, the hyperfine
interaction induces polarization in the $^3$He.  The resultant $^3$He
polarization saturates at a value
\begin{equation}
P_{\rm He}=\frac{\gamma_{\rm SE}}{\gamma_{\rm SE}+\Gamma}P_{\rm Rb},
\label{eq:relax}
\end{equation}
where $P_{\rm Rb}$ is the average Rb polarization produced by laser
optical pumping and $\gamma_{\rm SE}$ is the spin-exchange rate from
the Rb to the $^3$He nucleus. $\Gamma$ is the $^3$He spin
relaxation rate due to paramagnetic impurities on the cell wall
and in the gas~\cite{fit69,tim71,new93}.
Due to the slow rate of the spin exchange polarization process, high
$^3$He polarization requires $^3$He relaxation rates $1/\Gamma \approx  
50-100$ hours \cite{fit69}.  Rb densities of $10^{14}-10^{15}$ cm$^{-3}$
are used.

Typical alkali-spin exchange apparatus can be found in
Refs~\cite{cum95,lar91,joh95}.
The most common lasers in use for the Rb optical
pumping in high density $^{3}$He targets are Ti:sapphire lasers and
high-power laser diode arrays (LDA) \cite{cum95}.

Two fiber-coupled LDAs \cite{opt97} were used in this
experiment.  The unpolarized laser beam from the fiber was first focused
and then linearly polarized with a beam polarizing cube. The linearly
polarized component, which traversed the cube, was first circularly
polarized with a quarter wave plate
and then refocused so that the sample cell was completely
covered by the light spot. The linearly polarized component which
reflected from the cube was redirected parallel to the first beam 
with a mirror, and then circularly polarized by a second
polarizing-cube/quarter-wave-plate pair.  Thus the cylindrical 10-cm long
cell was covered with four circularly polarized round light spots.
The wavelength of the lasers were monitored by a
spectrometer~\cite{ocean}.
\subsubsection{$^3$He cells}
The state of the art of the glass cell preparation for high $^3$He
polarization is discussed in Ref.~\cite{chu95}.  The cell construction is
designed to optimize the FOM of the spin filter and meet practical
considerations as well.  The cells of this experiment were 10 cm long and
3.5 cm in diameter.  The cylindrical part
of the cell was produced from reblown Corning 1720 aluminosilicate glass
with a wall thickness of about 3 mm. The ends of the cell, the neutron beam
windows were made from 3~mm thick flat disks using $^{10}$B-free and
low-iron content Corning 1720 glass. $^{10}$B has a large neutron capture
cross section at thermal energies and would produce significant beam
attenuation if present in the windows.  Six cells were prepared for the
experiment. The
cells were loaded with 67 mbar of N$_2$ and 3 -- 11~bar $^3$He
corresponding to $^3$He densities of $1 \times 10^{21} - 2.7 \times
10^{21}$~cm$^{-2}$.  A destructive
pressure test with water was performed with a few cells which had the same
design.  The maximum pressure before the burst was measured to be about
16~bar.  The $^3$He polarization time constants varied from 40~h to 60~h.
The cell used for this measurement had 3.3~bar $^3$He, for a $^3$He
density of $8.4 \times 10^{20}$~cm$^{-2}$, and a measured relaxation
time of approximately
45 hours.  The highest polarization $^3$He with this cell was measured
using the neutron beam to be 45\%.  Due to technical difficulties, this
polarization could not be maintained, and the average polarization
achieved for the measurement was near 20\%.
The $^3$He polarization direction was
perpendicular to the neutron beam. The 86-cm diameter coils in the
Helmholtz configuration were used to provide the
holding field of 3.0 mT. The $^3$He polarization was monitored by
the adiabatic fast passage method \cite{cou88,cou90}.

The oven required to control the Rb density in the sample
cell contained two almost identical $^3$He cells. One of the cells
was continuously polarized by two lasers: the second cell, which contained
approximately the same amount of $^3$He as the first, did not have any Rb
so the $^3$He nuclei were not polarized.
This cell was used as a reference cell. The cells were kept at temperature
of about 175~$^{\rm o}$C for the optimum Rb density.
The whole $^{3}$He apparatus was mounted
on a movable table which allowed the cells to be alternated into and out of 
the beam every two minutes by switching the position of the table whose
position was controlled with an accuracy of 0.2 mm.
In one table position 2400 neutron pulses were accumulated, after which
the table was moved to the other position.  Each two-position cycle
comprises one run.
The data acquisition system separated information between the polarized
cell and reference cell.
\subsection{$^3$He Scintillation Counter}
A $^3$He scintillation detector was chosen as the transmission
detector for its high efficiency in this neutron energy
range, fast time response ($\approx 100$ nsec), and 
insensitivity to gamma rays. The $^{3}$He scintillation detector utilizes
the same large capture cross 
section for neutrons on $^{3}$He that is used in the polarizer: 
$n+{^3}He \rightarrow p+{^3}H+767~keV$. Because the capture cross
section
depends on the neutron energy, the $^3$He thickness in the detector
can be selected so that it has the optimum stopping power for the
neutrons of interest.  Thus the neutron detection efficiency of the cell
with a given pressure depends on the neutron energy.  However, since the
neutron kinetic energy is small compared to the $Q$ value of the reaction, 
the pulse height of the neutron capture event by $^3$He is constant.

\begin{figure}
\scalebox{1.2}{
\includegraphics*[1.75in,4.25in][8in,8in]{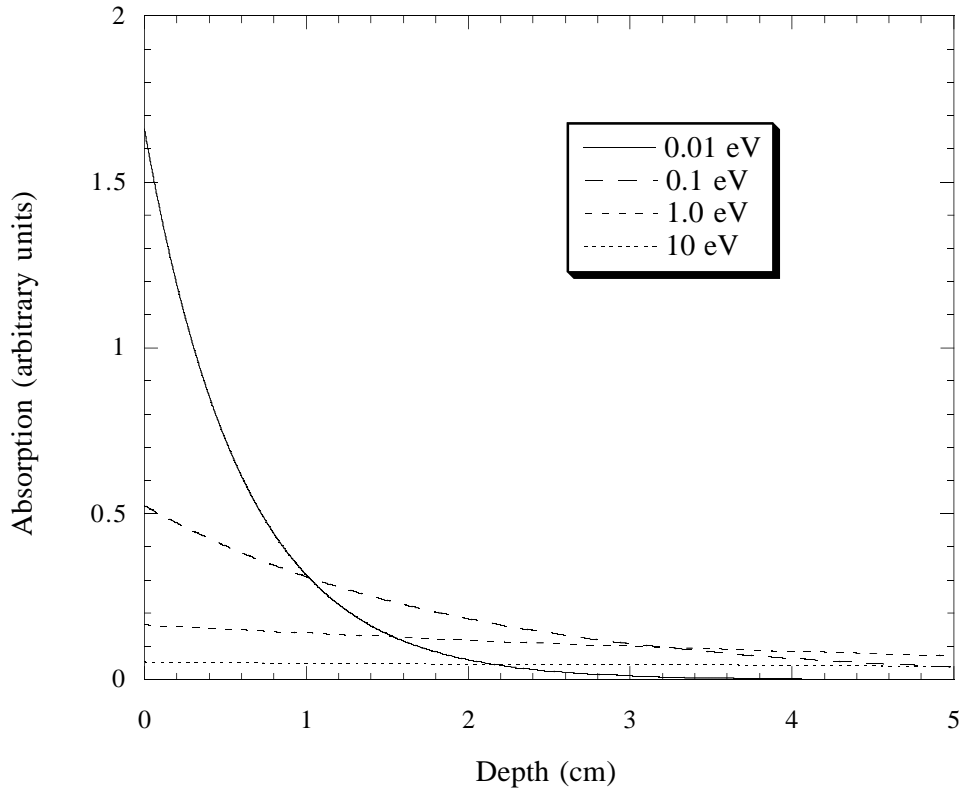}}
\caption{Transmission profile for neutrons in the $^3$He scintillator
detectors.  As a function of neutron energy, it can be determined that the
average transmission depth in the detector is given by
equation~\ref{eq:dave}.}
\label{fig:profile}
\end{figure}

The products from the reaction, triton and proton,
have kinetic energy of 575~keV and 192~keV, respectively.
The triton has a mean free path of
0.9~bar-cm in $^{3}$He gas and the proton mean-free path is
5.83~bar-cm.  In the experiment we used a single cylindrical $^3$He
scintillation detector cell with a diameter of 5 cm and length of 5 cm.
We used typically 8~bar of $^3$He in the detector cell.
The great majority of capture events deposit all their
energy in the detector.  The detector efficiency is then given by the
neutron absorption in the cell:
\begin{equation}
e = 1 - T^0_n = 1 - exp(-nl\sigma_{re})
\end{equation}
A small correction must be made to the detector efficiency when neutrons
are captured near the wall of the detector, and subsequently do not deposit
all their energy within the detector.  This effect is energy
dependent, since low energy neutrons have a higher absorption cross
section and are therefore captured nearer the front of the
detector (see figure~\ref{fig:profile}).  Since we require only a
transmission $ratio$, and the efficiency and ``wall effect'' are
spin-independent, for the neutron polarization measurements no corrections
need be made to the raw data to account for these effects.  However, the
energy dependence of the efficiency has a small effect on the calibration
of the neutron TOF data in terms of neutron energy (discussed in section
4.2).

The detector cell was made from aluminum with a 5 mm thick
entrance window for neutrons at the end of the cylinder.
Scintillation light was detected through a 3 mm thick sapphire window
which was mounted on the other end of the cylinder and was sealed with a
Viton O-ring.  The photomultiplier tube was a 2 inch diameter Amperex
XP2262B which provided a fast rise time ($2.0$ nsec) and a good quantum
efficiency ($28$\% at $400$ nm).

The two charged particles, proton and triton,
from the capture reaction produce vacuum ultraviolet (VUV)
scintillation light in $^3$He gas through
de-excitations of short-lived $^3$He quasi-molecules \cite{gri87}.
The n-$^3$He reaction is estimated to produce
$\approx$ 1-2 VUV photons per keV in $^3$He gas.
It was impractical in our case to efficiently detect VUV photons.
Therefore, the inside 
walls of the cell were coated with a wavelength shifter (WLS) to
convert the VUV light to the
visible region. The WLS was tetraphenylbutadiene (TPB, 
1,1,4,4-Tetraphenyl-1,3-butadiene; 
Fluka $88020$).  The fluorescence
 efficiency of TPB has been studied recently by McKinsey {\it et al.} 
\cite{kin97}.
 After the coatings were made, 
the cell was held at a temperature
of $60-70^o$~C and pumped to a pressure below $1.3\times10^{-6}$ mbar for
a day.

\begin{figure}
\scalebox{.8}{
\includegraphics[1in,1.5in][8in,9in]{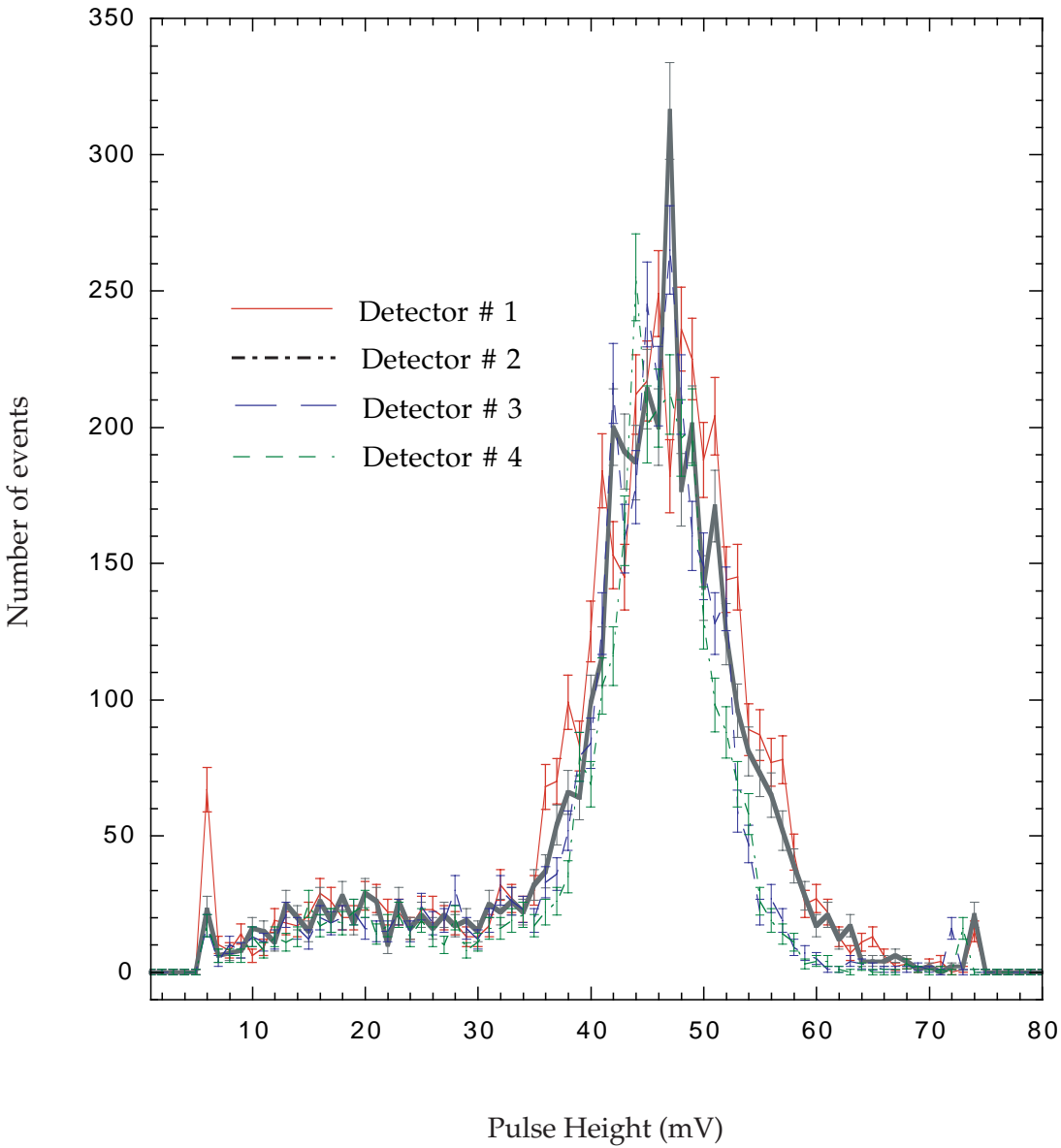}}
\caption
{
Pulse height distribution from four $^3$He scintillation detectors used in
this experiment.  The spectra are measured for 2~eV neutrons.
}
\label{fig:phd}
\end{figure}

To increase the light output and improve the pulse height resolution,
7\% of Xe gas by volume was added \cite{nor58}.
Figure~\ref{fig:phd} shows the pulse height distribution with the
$^3$He-Xe mixture measured with thermal neutrons. 

Pulses from the photomultiplier tube 
	were fanned into three separate integral level
	discriminators whose outputs are in turn recorded
	in scalers.  The first of the discriminator-scaler pairs is used to
	determine the neutron yield from which the transmission ratio
	is extracted, 
	and is described in more detail in Sec. 3.5.
	The remaining two discriminator-scaler pairs are used
	to stabilize the gain of the photomultiplier tube.
	These discriminators are set at 60~mV and 120~mV, 
	respectively.  The lower discriminator is set in the minimum of 
the pulse height distribution from the detector and counts essentially all
of the neutron events. The upper discriminator is centered at the peak of
the pulse height distribution and has a count rate about half that of the 
lower level discriminator.  
At approximately five minute intervals, the data acquisition system reads both 
scaler values and adjusts the voltage of the photomultiplier
tube until the scaler value associated with the 
60~mV discriminator is approximately twice that of the
120~mV one.  By this method the gain of the PMT was stabilized to the 
1\% level.  Given the spectrum of pulse height shapes and the location of 
the discriminator windows, this stability 
suffices to reduce the fluctuation of the number of recorded neutron counts 
due to detector gain changes to a value which was below counting
statistics.
\subsection{Data Acquisition System}
Pulses from the $^3$He scintillator are first passed through a 100
ns filter for shaping.  They are then passed through a 10 $\times$
amplifier before being pulse-height discriminated and
	recorded in an Ortec T914~\cite{egg} multiscaler
	that is divided into 9625 bins.  Individual neutron counts are
	accumulated in
	a single multiscaler bin for a preset length of time, 
	$t_{\rm dwell}$, until an external oscillator
	signals the multiscaler to advance to the next bin.
	The data are thus stored as histograms representing
	the number of scaler counts versus the neutron time of flight.

	When the last TOF bin has been reached, the
	multiscaler returns to the first bin and waits for
	the next beam burst, which is preceded by a start 
	signal from the proton storage ring.  The same start
	signal triggers a countdown scaler set at
	2400.  New counts in each TOF bin are accumulated
	on top of those from the previous beam burst
	until the countdown scaler reaches zero.  At this time data
	acquisition is halted, and the data stored in the
	multiscaler is written to disk.  This process requires
	approximately one minute, during which the translation
	table for the target and empty cells
	is moved to the next position.
	The countdown scaler is reset to 2400, the multiscaler
	is cleared, and data acquisition recommences only after
	the multiscaler is finished writing to
	disk and the position transducer indicates
	that the translation table is in the correct
	position.  Each target-in, target-out cycle
	comprises one run.  

The multiscaler is read into CAMAC through a Kinetic Systems model 3344
serial line communication interface.  The data-acquisition is based in
VME, with online sorting handled on an HP-workstation by the XSYS
data-acquisition package, as developed at IUCF \cite{yod93}.
\section{Data Analysis and Results}
In this section we describe the details of the measurement strategy, the
data analysis, and the results.
\subsection{Accurate neutron beam polarization measurement with transmission}
\label{sec:measurement}
The following method, which is a slight 
modification of the technique outlined earlier (defined by
equation~\ref{eq:polVratio}), was used to measure 
the neutron polarization.  Both a target cell and a reference cell were
placed on a translating table.  Transmission through the target cell
(polarized and unpolarized) was normalized by the transmission through the
unpolarized reference cell.  The relative normalization was measured every
six minutes, thus reducing the effects of time-dependent drifts in the
properties of various parts of the system to negligible levels.

\begin{figure}
\scalebox{1.2}{
\includegraphics*[1.75in,4in][7.5in,8in]{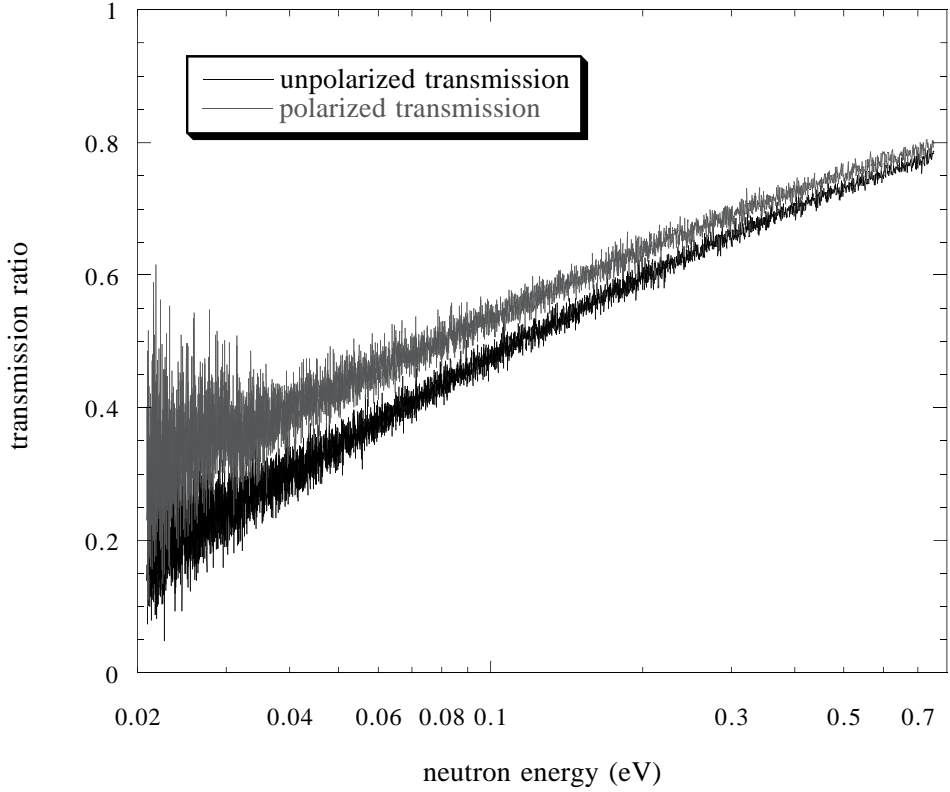}}
\caption
{
Online indication of the transmission enhancement observed in this
measurement, caused by polarized $^3$He.  Shown are the transmission
ratios $T_{n}/T^{ref}_{n}$ for the target cell unpolarized
(lower curve) and polarized (upper curve).
}
\label{fig:enhance}
\end{figure}

To determine the neutron beam polarization, we first obtained the ratio of
the transmissions through the {\bf unpolarized} target
cell ($T^0_n$) and the reference cell ($T_n^{ref,1}$):
\begin{equation}
K=\frac{T_{\rm n}^{\rm 0}}{T_{\rm n}^{\rm ref,1}}.
\label{eq:norm2}
\end{equation}
The factor $K$ is energy dependent, and dominated by the differences in
the $^3$He and window thicknesses for the two cells.
We then polarized the target cell, and obtained the ratio of the
transmissions through the target cell while {\bf polarized}, $T_{\rm n}$,
and the reference cell (always unpolarized), $T_{\rm n}^{\rm ref,2}$. From
equation~\ref{eq:trans2} we obtain the ratio of these transmissions
\begin{equation}
\frac{T_{\rm n}}{T_{\rm n}^{\rm ref,2}}=
K\cosh{(n \sigma_{p} lP_{\rm He})}.
\label{eq:norm1}
\end{equation}
The values of both $K$ and $\cosh(n \sigma_{p} lP_{\rm He})$ depend upon
the neutron energy. 
As long as the properties of the reference cell are static ({\em i.e.},
constant $^3$He density), $T_{\rm n}^{\rm 
ref,1}=T_{\rm n}^{\rm ref,2}$ and we can take the ratio of
equations~\ref{eq:norm1} and \ref{eq:norm2} to get the ratio $T_{\rm
n}/T_{\rm n}^{\rm 0}$ needed in equation \ref{eq:polVratio}
for extracting the neutron polarization:
\begin{equation}
\frac{T_{\rm n}}{T_{\rm n}^{\rm 0}}=
\cosh{(n\sigma_{p}lP_{\rm He})}
\label{eq:norm3}
\end{equation}
We thereby determine the argument, $n \sigma_{p} lP_{\rm He}$, that can
be substituted into equation~\ref{eq:pol1} to obtain beam polarization,
$P_{\rm n}$. None of the individual parameters in the argument need be
known.   Figure~\ref{fig:enhance} shows
the transmission enhancement caused by $^{3}$He polarization as a
function of neutron energy.
\subsection{Neutron Time-of-Flight (TOF) and Energy Calibration}
The neutron energy is determined by TOF, which for nonrelativistic
neutrons is inversely proportional to the square-root of the neutron
energy.  For optimal resolution a long flight path and fine time binning 
are important.  For this reason, and also to avoid a number of
systematic effects associated with non-forward neutron scattering from the 
$^3$He target, the neutron detector is located approximately 60~m away from
the neutron source and more than 50~m from the $^{3}$He target.
Calibration of the TOF in terms of neutron energy is determined by
installing thin films of $^{238}$U and natural Ir into the beam. These
nuclei possess neutron resonances whose absolute energies are known to
0.1\% or better (see table~\ref{tab:tofresults}).  Using the FITXS
routine mentioned earlier, the centroid channel number of each of these
resonances was extracted and correlated with the resonance energy.
FITXS takes into account details of the beamline geometry and moderator
properties including the emission time distribution of the neutrons from
the moderator as a function of TOF.
   \begin{table}
    \caption{Resonant nuclei used for the neutron time-of-
                        flight energy calibration.}
                \begin{tabular}{|c|c|}
\hline
                        Nucleus  & Resonance Energy (eV) \\
\hline
                        $^{191}$Ir     &   0.6528  $\pm$ .0005 \\
                        
                        $^{193}$Ir     &   1.302  $\pm$ .001 \\
                        
                        $^{238}$U     &   6.671  $\pm$ .002 \\
                        
                        $^{238}$U     &   20.872  $\pm$ .006 \\
                        
                        $^{238}$U     &   36.680  $\pm$ .011 \\
                        
                        $^{238}$U     &   66.02  $\pm$ .02 \\
                        
                        $^{238}$U     &   189.67  $\pm$ .04 \\
\hline
               \end{tabular}
               
               \label{tab:tofresults}
        \end{table}

The neutron motion is nonrelativistic to an excellent approximation and
the relation between the neutron kinetic energy and the TOF parameters is
given by
\begin{equation}
E_{n} =\frac{1}{2}M_{n}(\frac{L_{ave}}{i(t_{dwell}-t_{off})}){^2}
\end{equation}
where $M_{\rm n}$ is the neutron mass, $L_{ave}$ is the average path length
from the source to the detector, $i$ is the channel number of the Ortec
T914 multiscaler, $t_{\rm dwell}$ is the size (in $\mu$s) allocated for
each TOF channel, and $t_{off}$ is a timing offset associated with the
electronic start signal.  Fitting the known energies of these resonances
to the centroid of the channel number at which they appear in the neutron
TOF spectra, we extract the TOF parameters $L_{ave}$ and $t_{off}$ for
neutrons of energies between 0.65 and 190 eV.  

\begin{figure}
\scalebox{1.2}{
\includegraphics*[1.85in,5.25in][7.5in,9in]{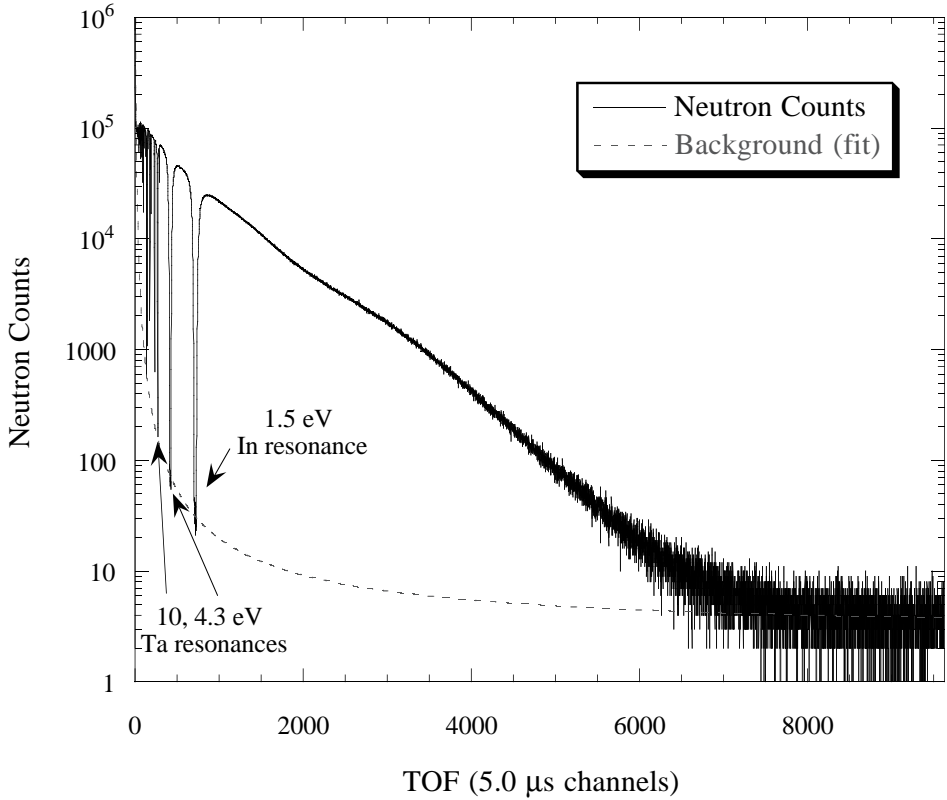}}
\caption
{
A time-of-flight spectrum after summing over 2$\times$10$^5$ neutron
pulses. The dips in the spectrum are caused by the resonances of the In
and Ta absorbers. The lack of neutrons at lowest energies (long TOF) is
a consequence of the absorption in the $^{3}$He.  The dashed line is a fit
to the total background.
}
\label{fig:TOF}
\end{figure}

The offset is dependent on the dwell time of each TOF channel.  For the
5.0 $\mu$s channels
used in this measurement, an offset of $t_{off}$ = 8.72(3) $\mu$s was
determined from the fit.  This uncertainty, which contributes at levels
below 0.1\% starting at channel $i = 30$ while our analysis begins at
channel 300 (see figure~\ref{fig:TOF}), makes negligible contributions to
the final uncertainty.

\begin{figure}
\scalebox{1.2}{
\includegraphics*[1.5in,4.25in][7.5in,8in]{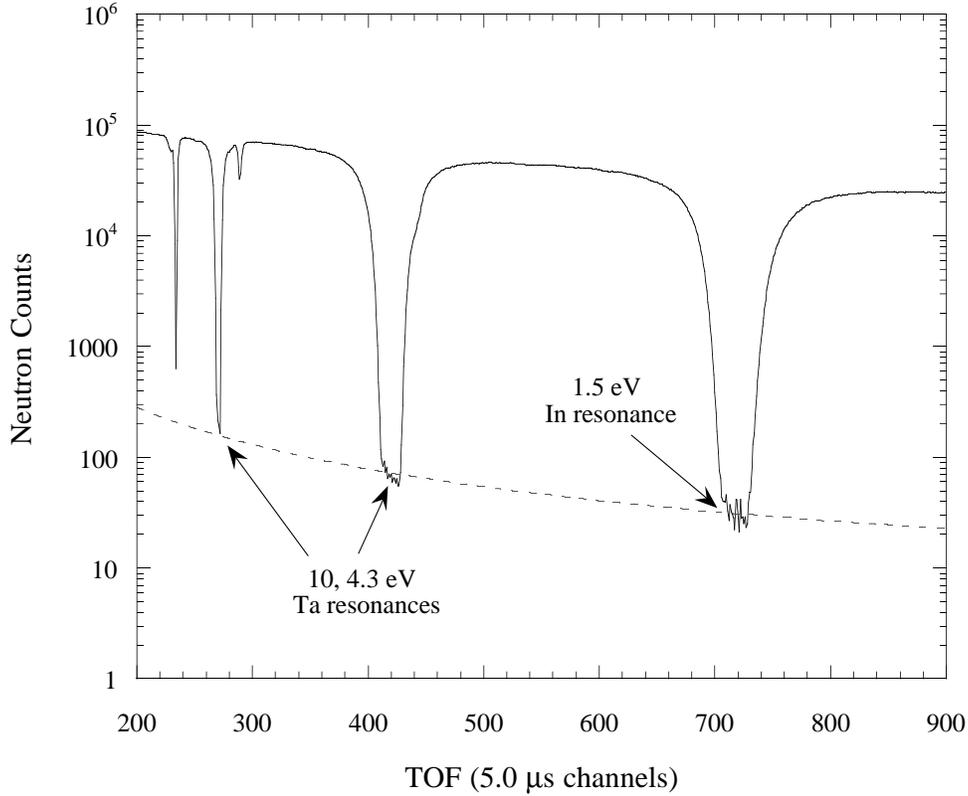}}
\caption
{
Same conditions as figure~\ref{fig:TOF}, zooming in on the resonances
used for the background fit.
}
\label{fig:TOFzoom}
\end{figure}

The average path length extracted was $L_{ave}$ = 59.752(1) m, a value
dominated by the high energy resonances used for TOF calibration.  However, 
a correction to the path length must be made due to the energy
dependent neutron penetration depth in the detector, which contributes
strongly at energies lower than those used in the calibration.  The
probability that a neutron will be absorbed at a given depth $d$ in the
detector is proportional to the neutron transmission to that depth, defined
as $T^0_n$ = exp($-nd\sigma_{re}$).  
A series of absorption profiles are shown in figure~\ref{fig:profile}.  
It may be shown (using equation~\ref{eq:trans2}, and the
$1/v$ dependence of the neutron capture cross section as discussed in
section 2) that the average penetration depth as a function of energy is
given by
\begin{equation}
d_{ave}=-\sqrt{E_n}/(\sigma_{0}n)\ln(1/2(1 
+ \exp(-\sigma_{0}nt/\sqrt{E_n})))
\label{eq:dave}
\end{equation}
where $n$ is the number density of $^3$He atoms in the detector, and $t$ is
the total thickness of the detector.  This correction is largest at low
energies ($40~meV~< E_n < 0.1~eV$), where it approaches 2 cm, and nearly
negligible at energies above those used in the fitting routines, 
($E_n > 0.65~eV$).  The uncertainty in this correction is due to the
uncertainty in
the precise knowledge of the detector pressure (of order 5\%), and to the
variance in the penetration depth given by:
\begin{equation}
\frac{\Delta{L}}{L} = \sqrt{\frac{<d^2> - <d>^2}{(L_{ave}-t/2)^2}}
\end{equation}
The net uncertainty in the path length is of order 0.04\% at high energies
(1 -- 10 eV) and 0.01\% at lower energies (0.01 -- 0.1 eV).  The net
uncertainty in the energy scale is then 0.08\% at high energies and 0.02\%
at lower energies (see figure~\ref{fig:error}).
\subsection{Backgrounds}
The largest source of backgrounds in this measurement are from fast
neutrons scattered and moderated in the beam pipe and neutrons produced in
secondary reactions at the source. These neutrons can reach the detector
at a time different from their equal energy counterparts and thus blur the
simple relation between neutron kinetic energy and TOF.  The first source
was greatly reduced by modifications to the collimation and shielding
described previously.  A lead wall was introduced to reduce the number of
fast neutrons that enter the evacuated beam pipe.  Within the beam pipe,
the collimation is entirely $^{10}$B and $^6$Li doped polyethylene, which
moderates and absorbs the extraneous neutrons rather than scattering them and 
forming 
secondary sources themselves.  Other sources of backgrounds are small by 
comparison.  Due to its low atomic number, the $^3$He scintillator is 
relatively insensitive to gammas present in the neutron beam.  Cosmic rays
background rates are low compared to the 20 kHz neutron counting rates. 

A typical TOF spectrum for this experiment is shown in
Figure~\ref{fig:TOF}.
Neutron resonances cause the dips in the transmission spectra.  The
resonances in this figure are from In and Ta foils inserted in the beam
(Table~\ref{tab:black}), chosen because they possess
large compound nuclear resonances at precisely known neutron energies.  The 
thicknesses of the In and Ta samples were chosen to make them opaque to
neutrons at these resonance energies.  Consequently, any counts observed
beneath one of these dips are either background counts or neutrons of
slightly different energy which arrive at the resonance energy TOF due
primarily to the spread of emission times from the moderator. The
background rates were measured
underneath these black resonances and at long TOF, where the low energy
neutrons are absorbed by the $^3$He in the beam.  The rates are then
normalized to the neutron beam flux as measured by the beam monitor.
Except for small TOF which is outside the region of our analysis, the
background rates are well described by the expression:
\begin{equation}  
                \tilde{N}_{\rm bkgnd} = \sum^3_{i=0} a_i/t^i,
                \label{bckgnd}
\end{equation}
where $\tilde{N}$$_{\rm bkgnd}$ is the measured background per channel,
and $t$ is the channel number (see figure \ref{fig:TOF}).  The background
is energy dependent, ranging from 0.2\% at 10~eV down to 0.06\% at 1.0~eV
and back up to 1.5\% at 40~meV.  The extracted fits of the backgrounds
to the above expression are accurate to better than 7\%, with the
subsequent uncertainty in the transmission maximal at the low and
high energy extremes (0.1\% at 40~meV, 0.015\% at 10~eV) and nearly
negligible in between (0.005\% at 1.0~eV) (see figure~\ref{fig:error}).
      \begin{table}
      \caption{Resonant nuclei used for the TOF background
                        measurements.  Only resonances opaque to the
                        neutron beam are listed.}
                \begin{tabular}{|c|c|}
\hline
                        Nucleus  & Resonance Energy (eV) \\
\hline
                        $^{115}$In   &    1.5 \\
                        
                        $^{181}$Ta   &    4.3 \\
                        
                        $^{181}$Ta   &   10.4 \\
                        
                        $^{181}$Ta   &   40 \\
                        
                        $^{181}$Ta   &   36 \\
                        
                        $^{59}$Co    &   120 \\
                        
                        $^{55}$Mn    &  340   \\
\hline
                \end{tabular}
                
                \label{tab:black}
        \end{table}
\subsection{Deadtime and Pulse Pileup}
If two or more events in the $^3$He scintillator with energies
below the discriminator threshold overlap in time, the
discriminator may register a count despite the fact that both events  
occur beneath the discriminator threshold.  We call this
pulse pileup:  it increases the observed count rate relative to the true
neutron rate at the detector.  In addition, the
detector-electronics combination is rendered insensitive for a brief time
after an event is registered.  We call this deadtime:  it 
decreases the observed count rate relative to the true neutron rate.

Pulse pileup and deadtime effects are difficult to model reliably.  We
therefore decided to measure the net effects of deadtime and pulse pileup
experimentally.  The first aspect of our measurement strategy is to keep
the deadtime and pileup corrections small.  The $^3$He detectors have a
response time of approximately 100ns.  Instantaneous counting rates at the
detector were intentionally reduced by the beam collimation to less than
50kHz (20 kHz with $^3$He in the beam).  An order of magnitude estimate of
the deadtime losses is then $\sim$0.2\%.  

The next aspect of our strategy is to ensure that the deadtime/pileup
corrections are dominated by processes occuring in the detector and not
later on in the electronics chain.  The ORTEC multiscaler possessed a 7
nsec deadtime, and the pulse width from the discriminator was 20 nsec.
Both of these values are smaller than the width of the pulses
from the detector, and therefore none of the electronics at or after the
discriminator contribute to the deadtime.

To first order, the observed yield $Y$ at the detector is then given by:
\begin{equation}
Y = R(1 - R\tau)
\label{eq:dt1}
\end{equation}
where $Y$(Hz) is the observed rate, $R$(Hz) is the true rate, and
$\tau$(sec) is a parameter which includes the effects of both deadtime
and pileup.  This relation is good for $R\tau$ $\ll$ 1, which our system
satisfies.

To determine this parameter $\tau$, we performed a measurement of the
total cross section on a nucleus whose neutron cross section is
independent of energy.  Measurements of the neutron total 
cross section may be performed with our apparatus by replacing the two
$^{3}$He cells with the sample. The cross section is given in terms of
measured parameters by:
\begin{equation}
\sigma = \frac{1}{nl}\ln(\frac{R_{out}}{R_{in}})
\label{eq:xs}
\end{equation}
where $\sigma$ is the total neutron-target cross section, $n$ is the number 
density of atoms in the target, $l$ is the target thickness, and
$R$$_{out}$ and $R$$_{in}$ are the true rates for the target out
and target in the beam respectively.  Using only the approximation that
the deadtime correction is small, these equations may
be combined to extract the deadtime parameter, $\tau$:
\begin{equation}
\sigma_{meas} = \sigma_{true} - \frac{\Delta R}{nl}\tau
\label{eq:dt2}
\end{equation}
where $\sigma_{meas}$ is the uncorrected cross-section with no deadtime
corrections applied to the transmission data (background corrections are
included), $\sigma_{true}$ is the true cross section of the sample,
and $\Delta$$R$ is the sample in versus sample out difference in rates (Hz).  

Total cross section measurements on a single carbon sample at four
different detector counting rates from approximately 20kHz to 120kHz were
performed.  Carbon was chosen because it has a total cross section which
has been accurately measured and is independent of neutron energy to an
excellent approximation in this energy range~\cite{sch88}.  The rates were
changed by varying the
thicknesses of Pb attenuators and In and Ta filters in the beam. The
extracted cross sections for the four measurements were then compared over
nine different energy regimes. The measured cross sections were plotted
versus the difference in rates to  extract the deadtime parameter from the
slope of the curve. This measurement of the deadtime parameter yields $\tau$
= 47(7) ns, which is a reasonable value given the 100 ns response of the
detectors coupled with pileup effects.

The instantaneous rates at the detector was 20 kHz at higher neutron
energies ($1.0 \leq E_n \leq 10.0 eV$), and near 200 -- 500 Hz at lower
neutron energies (0.4 -- 0.5 eV).  The correction to the detector yield is
then $\sim$0.1\% at the higher neutron energy, 0.01 -- 0.025\% at
lower neutron energy.  We exploit the fact that, since we perform a
relative transmission measurement, we are sensitive ultimately only to the
difference in deatime/pileup corrections for the polarized and unpolarized
$^3$He --- the correction to the transmission ratio is proportional
to $\tau$$\Delta$R.  The difference in thicknesses of the reference and
target cells was $\sim$25\%, making the correction to the transmission
ratio 0.020(4)\% for $1.0 \leq E_n \leq 10.0 eV$, and
$\leq$ 0.0025(4) for $E_n \leq 0.5 eV$ (see figure~\ref{fig:error}). 
\subsection{Results}
We constructed a model based on the operation of the neutron polarizer as 
outlined in the introduction and on the behavior of the neutron cross
section on $^{3}$He discussed in Section 2 and used it to fit our data. 
211 runs were accumulated with the target cell unpolarized, 119 with the
target polarized.  Additional runs were accumulated while the target was
being polarized.  The rise of the $^3$He polarization is very nearly an
exponential, with possible deviations due to fluctuations in Rb density
(from temperature fluctuations), for example.  By observing the change in
transmission as the $^3$He polarization slowly increased, we extracted
the rise time of the polarization, $\tau_{rise} = 6.0
\pm 0.2$ hours, and the asymptotic value of the transmission parameter
$nl\sigma_{p}P_{He}$.  The 119 ``polarization'' runs were accumulated with
the $^3$He initially within 0.8\% of the asymptotic transmission, with later
runs within 0.1\%.  For each run the transmission through each $^3$He cell
was normalized to the beam monitor, and corrected for background and
deadtime/pileup effects.  The transmission through the polarized $^3$He cell
was then scaled at each run by a factor:
\begin{equation}
F = \cosh(nl\sigma_{p}P_{He,asym})/\cosh(nl\sigma_{p}P_{He})
\end{equation}
from equation~\ref{eq:trans2}, where $P_{He,asym}$ is the asymptotic
value of the $^3$He polarization, and the polarization in the denominator 
is determined by the time constant noted above.  The runs
were then summed together to improve the statistics for analysis.  The
effect of the scaling was negligible at high energies (for $E_n >
1.0~eV$, the average effect was $0.0 \leq F - 1 \leq 0.002(3)\%$) and
largest at low energies (for $E_n \approx 40~meV$, $F - 1 \approx
0.40(3)\%$).  See figure~\ref{fig:error}.  

Following the experimental procedure outlined in
section~\ref{sec:measurement}, we first consider the {\bf unpolarized} target
runs. The transmission ratio for both target and reference cells
unpolarized is given (from equation~\ref{eq:trans2}) by:
\begin{equation}
K=\frac{T^0_{n}}{T^{ref,1}_{n}}=\exp(\Delta(nl)\sigma_{re}+C)
\end{equation}
where $\Delta(nl)$ is the difference in the $^3$He thicknesses, and the
coefficient $C$ allows for an energy independent difference in
transmission  (due to differences in potential scattering from the windows
of the glass cells, for example).  Since the cross section follows the
well known $1/v$ dependence described earlier, this relation can be
rewritten as a function of TOF:
\begin{equation}
K=\exp[\Delta(nl)\sqrt{\frac{2E_{0}}{M_n}}\frac{\sigma_0}{L}(t 
-t_{off}) +C]
\end{equation}
where $\sigma_{0}$ is the cross section at the neutron energy $E_{0}$,  
$M_{n}$ is the neutron mass, $L$ is the flight path length, $t$ is the
neutron time of flight, $t_{off}$ is the offset between the start of the beam 
burst and the start of counting at the data acquisition system.  A fit to the
transmission data for both cells unpolarized yields a measured difference
in target and reference cell thicknesses of
$\Delta(nl)$~=~2.896(3)~$barn^{-1}$
(consistent with a target cell thickness of $\sim$33 bar-cm and a
reference cell thickness of $\sim$40 bar-cm). $A$ was measured to be
-0.357(4). This means that the difference in neutron transmission
through the two cells at very
high neutron energies was about 30\%.

The transmission ratio for runs with {\bf polarized} target and unpolarized
reference cell was then divided TOF-channel-by-channel by the unpolarized
target ratio, yielding the normalized transmission ratio in 
equation \ref{eq:norm3}.  Rewriting as a function of energy:
\begin{equation}
\frac{T_{\rm n}}{T_{\rm n}^{\rm 0}}=
\cosh{[nlP_{\rm He}\sigma_{0}\sqrt{{E_{0}/E_n}}}].
\end{equation}

\begin{figure}
\scalebox{1.2}{
\includegraphics*[1.65in,4.2in][7.5in,8in]{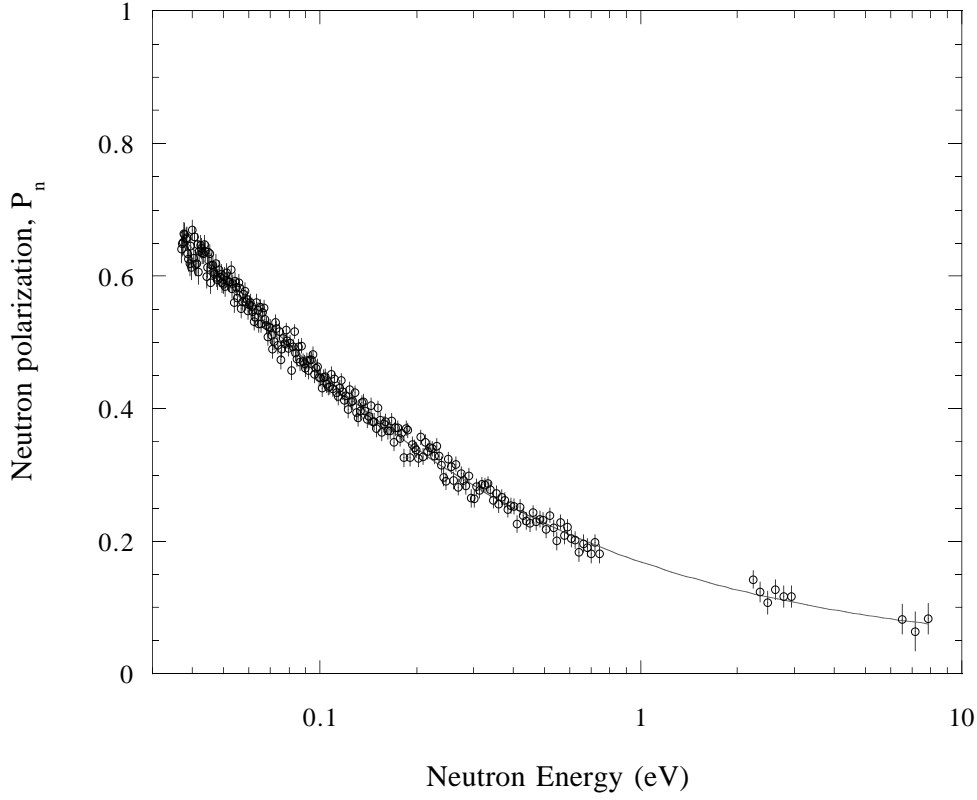}}
\caption
{
Measured polarization as a function of neutron energy.  The transmission
ratio is fit according to equation~\ref{eq:norm3}, then inserted into   
equation~\ref{eq:polVratio}.
}
\label{fig:fitpol}
\end{figure}

A fit of the transmission data to this functional form yields the product
$nlP_{He}\sigma_{0}\sqrt{E_{0}}=~.1583(4)~\sqrt{eV}$.  Presuming a
$\sim$30 
bar-cm target, this corresponds to a $^3$He polarization of $\sim$20\%.
The uncertainty in this parameter yields an energy dependent statistical
uncertainty that varies from 0.3\% at 40~meV to 0.01\% at 10~eV.
Systematic effects, dominated by uncertainties in backgrounds,
deadtime/pileup effects, energy calibration, and the time dependence of
the target polarization (see the following section) contribute an
additional energy dependent error, which peaks at high and low energies
near 0.1\%.  At a neutron energy of 0.1 eV, the neutron
polarization is 0.4616(15).  The polarization as a
function of energy given by this fit is shown with the data in
figure~\ref{fig:fitpol}.
\section{Statistical Accuracy and Systematic Effects}
\begin{figure}
\scalebox{1.2}{
\includegraphics*[1.75in,5.25in][7.5in,9in]{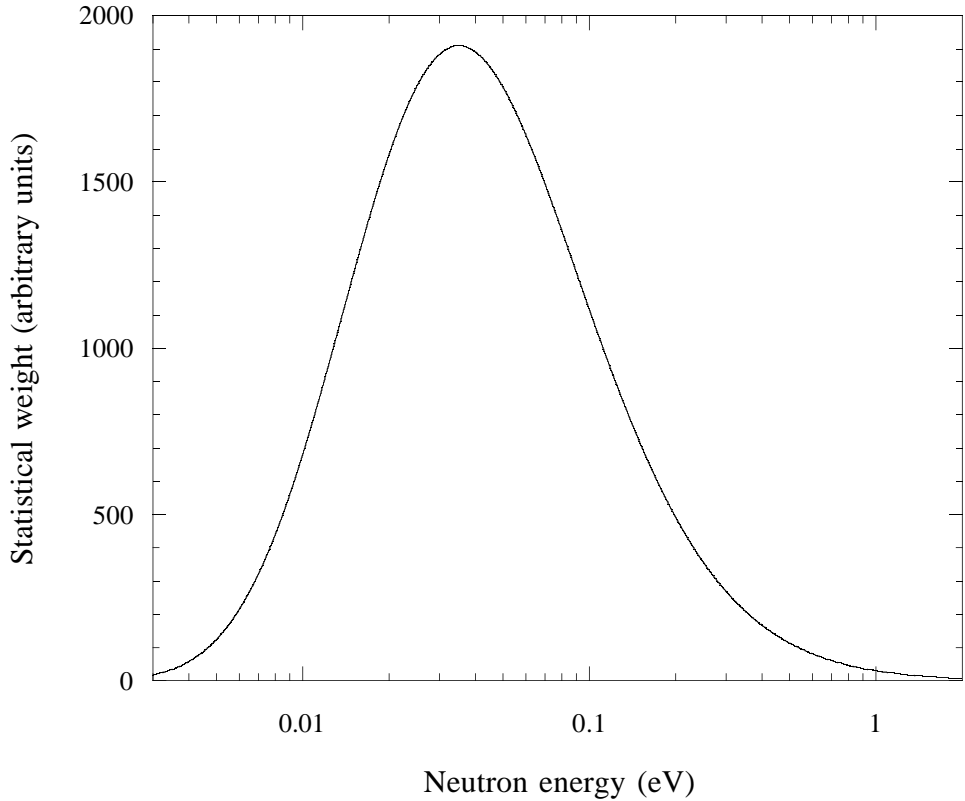}}
\caption
{
Statistical weight versus neutron energy.  The expected precision as
a function of energy range was approached in this measurement, as the
realized energy range (40 meV - 10 eV) overlapped significantly with the
idealized one for the particular target used (20 meV - 0.1 eV).
}
\label{fig:stat}
\end{figure}

The measurement was limited in accuracy by counting statistics.
Since the beam polarization measurement is a transmission measurement, the
statistical precision of the measurement can be estimated by calculating
the statistical weight over the neutron energy range of interest:
\begin{equation}
\sigma= \frac{1}{\sqrt{W}},
\end{equation}
where
\begin{equation}
W=t \int dE_n \Delta N \cdot P_{\rm n}^{2}(E_n)
\cdot T_{\rm n}(E_n) \cdot c(E_n)
\end{equation}
Here $t$ is the run time of the measurement, $dE_{n}$ is the energy bin,
$\Delta N$ is the number of neutrons per second with energies between
$E_{n}$ and $E_{n}+dE_{n}$ (corrected for the presence of background
filters In and Ta and other attenuating material in the beam) (see
equation~\ref{eq:flux}), $T_{n}(E_n)$ is
the neutron transmission from equation~\ref{eq:trans2}, $P_{n}(E_n)$ is
the neutron polarization from equation~\ref{eq:polVratio}, and $c(E_n)$ is
the detector efficiency.  In figure~\ref{fig:stat} we have plotted the
calculated weight as a function of neutron energy for a 3.3~bar-cm target 
cell with $P_{He}$ = 20\%.  The statistical precision of this measurement
calculated in this way is estimated to be 0.16\%.  Achievement of the
0.1\% absolute accuracy goal for this measurement, with
the given neutron flux, would require either 2.5 times the total run time
(from 33 hours to 82.5 hours) or an improvement in $^3$He polarization to
32\% or better.

There are a number of small systematic effects which must be considered to
assess the absolute accuracy of this technique. The
transmission formula gives the polarization of the neutrons under the    
following assumptions: (1) the rate of removal of
the polarized $^{3}$He atoms from the target by the beam is negligible,   
(2) the neutron beam is perfectly collimated and the neutron scattering
in the target is negligible in comparison with the absorption cross
section, so that all of the neutrons pass through the same length of
target, (3) multiple scattering into the
forward direction is negligible, (4) the neutrons are not depolarized as
they pass out of the $^{3}$He by spin-flip scattering in the window, (5)
the spin-dependent attenuation of the neutrons in $^{3}$He is the only
neutron spin-dependent process in the target, 
(6) changes in the kinetic energy of the neutrons as they pass into and
out of the holding magnetic field and the neutron optical potential for
the polarized $^{3}$He gas are negligible,  so that the neutron kinetic
energy in the target is given correctly by measurements outside the
target, (7) the $^{3}$He polarization in the target cell is uniform across
the beam profile, (8) the spin dependent Mott-Schwinger scattering can be 
neglected, (9) the effect of the finite temperature of the $^{3}$He gas is 
negligible, (10) the incident neutron polarization is zero.  Because of
the large distance between the target
and detector in this measurement (the detector intercepts a solid angle
of $2 \times 10^{-7}$ sr), almost all of these effects are completely 
negligible. Such effects in the
gas are further suppressed by the absorption cross section in $^{3}$He,  
which is large compared to the scattering and absorption cross sections
of window materials and the scattering cross section of $^{3}$He itself. 

\begin{figure}
\scalebox{1.2}{
\includegraphics*[1.6in,4.25in][7.5in,8in]{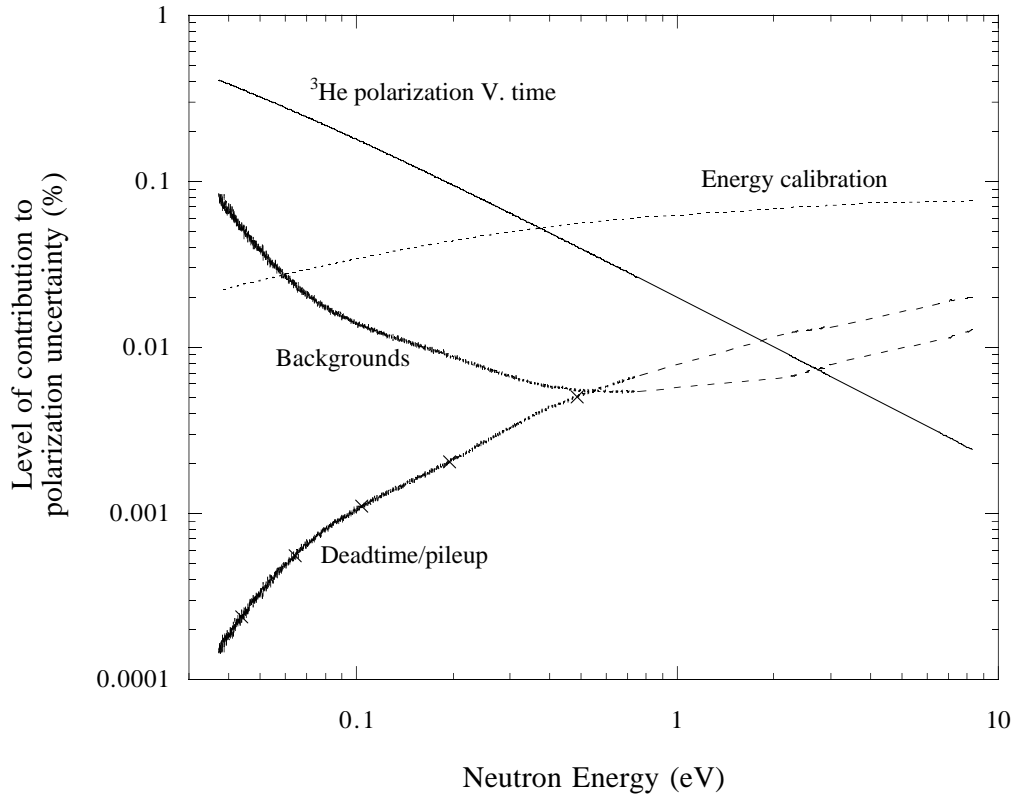}}
\caption
{
Systematic contributions to the final error in the neutron polarization
measurement, as a function of neutron energy.
}
\label{fig:error}
\end{figure}

\begin{table}
\caption{Systematic effects in the measurement of absolute neutron
polarization in this experiment.  The correction is listed from (40~meV
-- 10.0~eV), zero is listed if the effect is smaller
than 0.01\%.}

  \begin{tabular}{|c|c|l|}
\hline
    Correction~({\%}) & Uncertainty~({\%}) & Source of correction \\
\hline
     0.0 -- 0.26 & 0.0 -- 0.05 & time dependence of target $^{3}$He
polarization\\

     0.0 & 0.02 -- 0.08 & energy scale determination \\

     -(1.5 -- 0.06) & 0.1 -- 0.0 & detector backgrounds\\

     -(0.0 -- 0.025) & 0.0 -- 0.004 & detector dead time and pulse
pileup\\

     0.0 & 0.0 & distribution of neutron path lengths in target\\

     0.0 & 0.0 & multiple scattering in target \\

     0.0 & 0.0 & depolarization of neutrons in glass of the cell \\

     0.0 & 0.0 & small angle scattering in glass of cell\\

     0.0 & 0.0 & corrections from Mott-Schwinger scattering\\

     0.0 & 0.0 & corrections from thermal motion of the $^{3}$He atoms\\
\hline
     -1.5 -- 0.17 & 0.10 -- 0.10 & Net correction for systematic effects
\\
\hline
  \end{tabular}
\label{tab:polarization}
\end{table}
In Table~\ref{tab:polarization} we show a list of systematic effects from
various sources for the target conditions in our experiment. We have
included all known effects which modify the transmission expression by at
least 0.01\% in the energy range of our measurement.  The largest
systematic effect is due to the slowly rising $^3$He polarization, an
effect which may be made negligible if more time is available for ramping
up the $^3$He polarization.  Otherwise, deadtime/pileup corrections,
detector backgrounds and energy calibration give the largest effects,
which are at the 0.02\% level.  These effects are plotted as a function of
neutron energy in figure~\ref{fig:error}. For this reason we are confident
that it is possible to improve the absolute accuracy of this technique to
the 0.02\% level.

One question that arises is whether or not all of the techniques described
in this measurement can be extrapolated to the lower neutron energies of
interest in fundamental physics experiments. For example, it
becomes impractical to calibrate the neutron TOF spectrum  by
transmission through neutron resonances with absolutely known
resonance energies, since such resonances broaden and disappear as one 
approaches  neutron threshold. One alternative to this procedure which is 
well-suited 
to a transmission measurement is to use the sharp change in transmission
through polycrystalline solids at neutron energies corresponding to Bragg
scattering conditions. The discontinuity in the total cross section for
neutron energies above and below the Bragg condition is clearly visible in
a neutron TOF transmission spectrum and occurs at a well-defined neutron
energy. In addition, the pulse height resolution of the $^{3}$He
scintillator detector will worsen as the low energy neutrons are absorbed
close to the detector wall and the reaction products do not deposit the
full energy of the reaction into the gas.  The neutron detector efficiency
then develops an energy dependence which is influenced directly by the
detector geometry. This problem can be avoided by using a detector not
based on absorption in helium gas. The sensitivity of the detector to
gamma backgrounds must then be assessed.  Finally, in many cases it may
not be practical to locate the transmission detector so far from the
polarized target. In this case all of the systematic effects not important
for our geometry must be evaluated in detail to judge the limit to absolute 
accuracy.  
\section{Summary}
Using a laser-polarized $^3$He spin filter we have measured the absolute
neutron beam polarization from 40 meV to 10 eV with an absolute accuracy
of 0.3\%, where 0.2\% is from counting statistics, and 0.1\% is the
systematic error.  The measurement
was performed with $^3$He polarized to about 20\% with the Rb-spin
exchange method.  We have demonstrated that the simplicity of the
mechanism of neutron polarization
in polarized $^3$He gas allows for a new and reliable way of measuring
and monitoring on-line the absolute polarization of
a low energy neutron beam to better than 0.1\%.
The use of $^3$He spin filters to measure and monitor the beam polarization
does not require a spin flipper or analyzer. It depends only upon the
well defined relationship for neutron
transmission through polarized $^3$He gas. The $^3$He filter provides
a possibility for continuous online measurement and control of the neutron 
polarization. The extension of the technique to lower neutron energies is 
straightforward.

In addition to the fundamental neutron physics applications, the development 
of neutron polarizers and analyzers based on
$^3$He spin filters is important for neutron scattering. Neutron scattering 
studies in the thermal and epithermal energy range
have suffered for years from a lack of a convenient source of polarized
neutrons. With a $^3$He spin filter the useful
range can be expanded to epithermal neutron wavelengths.  

One of the advantages of the $^3$He spin
filter is its simplified experimental configuration.
One does not require any major modifications to the beamline for a beam
polarization measurement to be made. Compact configurations are possible
because there will be no gamma production and the $^3$He filter has a large 
acceptance angle, allowing the filter and experiment to be mounted close
to the source. The accurate absolute measurement described in this work
demonstrates that 
the simple relations between the polarized target 
parameters and the neutron transmission and polarization are satisfied to
high accuracy in practical polarized $^{3}$He devices.  

\section{Acknowledgements}

The authors would like to thank Mr. M. Souza for his enthusiasm, advice,
and great glass blowing skills, and Mr. T. Langston and Mr. J. Sandoval
for their technical help during the experiment.
This work has benefited from the use of the Los Alamos Neutron Science
Center at the Los Alamos National Laboratory. This facility is funded by
the US Department of Energy and operated by the University of California
 under Contract W-7405-ENG-36.  The work was
also supported by NSF CAREER Award NSF-PHY-9501312 and DOE award
DE-FG02-96ER45587.


\begin{thebibliography}{99}
\bibitem{egg} Model number is Turbo-MCS T914, EG\&G ORTEC.  No approval or
endorsement of any commercial product by the National
Institute of Standards and Technology is intended or implied. Certain
commercial equipment, instruments, or materials are identified in this
report in order to facilitate understanding. Such identification does
not imply recommendation or endorsement by the National Institute of
Standards and Technology, nor does it imply that the materials or
equipment identified are necessarily the best available for the
purpose.
\end{thebibliography}
\end{document}